\RequirePackage[OT1]{fontenc}
\documentclass{IEEEtran}
\IEEEoverridecommandlockouts
\usepackage{cite}
\usepackage{amsmath}
\usepackage{amsmath,amssymb,amsfonts}
\usepackage{multirow}
\usepackage{booktabs}
\usepackage{algorithmic}
\usepackage{graphicx}
\usepackage{textcomp}
\usepackage{xcolor}
\usepackage{float}
\usepackage{tabularx}
\usepackage{adjustbox}
\usepackage{hyperref}
\usepackage{url}
\usepackage{xurl}

\def\BibTeX{{\rm B\kern-.05em{\sc i\kern-.025em b}\kern-.08em
    T\kern-.1667em\lower.7ex\hbox{E}\kern-.125emX}}

\title{SRAM-SUC: Ultra-Low Latency Robust Digital PUF}

\author{Ayoub Mars, Hussam Ghandour, and Wael Adi}
\begin{document}

\maketitle
\begin{abstract} Secret Unknown Ciphers (SUC) have been proposed recently as digital clone-resistant functions overcoming some of Physical(ly) Unclonable Functions (PUF) downsides, mainly their inconsistency because of PUFs analog nature. In this paper, we propose a new practical mechanism for creating internally random ciphers in modern volatile and non-volatile SoC FPGAs, coined as SRAM-SUC. Each created random cipher inside a SoC FPGA constitutes a robust digital PUF. This work also presents a class of involutive SUCs, optimized for the targeted SoC FPGA architecture, as sample realization of the concept; it deploys a generated class of involutive 8-bit S-Boxes, that are selected randomly from a defined large set through an internal process inside the SoC FPGA. Hardware and software implementations show that the resulting SRAM-SUC has ultra-low latency compared to well-known PUF-based authentication mechanisms. SRAM-SUC requires only $2.88/0.72 \mu s$ to generate a response for a challenge at 50/200 MHz respectively. This makes SRAM-SUC a promising and appealing solution for Ultra-Reliable Low Latency Communication (URLLC). 
\end{abstract}


\section{Introduction}
\label{sec:introduction}
Cybersecurity is a primary concern in nowadays connected devices; IoT connected devices should be able to perform the functionality of their design in a secure way. This requires that each connected device has its unique clone-resistant or unclonable identity. Physical(ly) Unclonable Functions (PUFs) \cite{ref1} have emerged as a promising solution to authenticate IoT devices. However, PUFs are analog in nature requiring Fuzzy Extractors (FEs) or Helper Data Algorithms (HDAs) to stabilize their noisy responses \cite{ref2}, resulting with high hardware or software overhead and latency \cite{ref3}. To overcome these PUFs downsides, authors have proposed a new concept of digital clone-resistant functions coined as Secret Unknown Ciphers (SUC). SUC is an internally self-generated random secure cipher inside a chip. Due to the fact that SUC is digital in nature, it is robust during the lifetime of the electronic device. The creation process of SUC requires that each connected device embeds a System on Chip (SoC) FPGA. \par
Nowadays, SoC FPGAs are gaining popularity as accelerators; Recently, Xilinx launches the world's fastest data center and AI accelerator ALVEO cards \cite{ref4}. Meanwhile, Intel launched its Programmable Acceleration Cards (PAC) \cite{ref5}. These FPGAs-based acceleration cards increase tremendously the performance of industry-standard servers. SoC FPGAs will be widely used also in IoT devices such as for accelerating intelligent vision, automation in industry 4.0, vehicle-to-anything (V2X), etc. In addition to performance enhancement, SoC FPGAs can be deployed for security applications, such as building unique unclonable or clone-resistant device identity, fast encryption, decryption, and hashing, etc. For instance, Intel PAC 5005 is based on Stratix 10 SX integrating an SRAM PUF from Intrinsic ID, which is also used in Microsemi ‘S’ grade devices of SmartFusion2 and IGLOO2 as hardware block.\par

A mandatory security requirement in IoT is devices authentication; Because of the huge growth of connected devices, devices authentication with a trusted third party and device-to-device (D2D) authentication, which will be supported in 5G networks, are going to be a cornerstone in network communication performance, especially in the servers’ side. 5G networks are architected to support three services: enhanced Mobile BroadBand (eMBB), massive Machine Type Communication (mMTC), and Ultra-Reliable Low-Latency Communication (URLLC).  URLLC \cite{ref6} is a set of features designed to support latency-sensitive applications such as industrial internet, smart grids, remote surgery and intelligent transportation systems. These applications require tight security \cite{ref6}. URLLC has a target latency of 1-millisecond \cite{ref7}.
Hence, authentication should not be only strong from a security point of view, but also should be performed with a lowest possible latency. PUFs with FEs or HDAs can provide a secure authentication mechanism, but they have two main limitations: (1) small number of challenge-responses because PUFs are equivalent to hash functions and (2) high latency that makes PUF-based authentication impractical for many real-time applications. This work presents a digital clone-resistant function overcoming both limitations.
\par
\textbf{Contribution.} This work has three main contributions: \textcircled{1} we propose a new mechanism for creating SUCs as robust digital PUFs in volatile and non-volatile SoC FPGAs. \textcircled{2} a new SUC design is proposed based on deploying 8-bit S-Boxes that are generated randomly and internally, inside the SoC FPGA, by deploying a set of 4-bit S-Boxes. The resulting SUC is coined as SRAM-SUC. \textcircled{3} An accurate comparative analysis of SRAM-SUC with well-known authentication mechanisms is presented showing that SRAM-SUC has extremely better performance compared to PUFs with FEs or HDAs, it is lowering the latency of generating a response to a challenge to more than 40 000 times compared to Quiddikey IP in SmartFusion2 SoC FPGAs. Hence, SRAM-SUC can be embedded as robust digital PUF in volatile and non-volatile SoC FPGAs, that can be deployed in IoT connected devices or specific applications in URLLC such as V2X and commercial aviation.

\section{State of the art of clone-resistant primitives}
\begin{figure}
\includegraphics[scale=0.47]{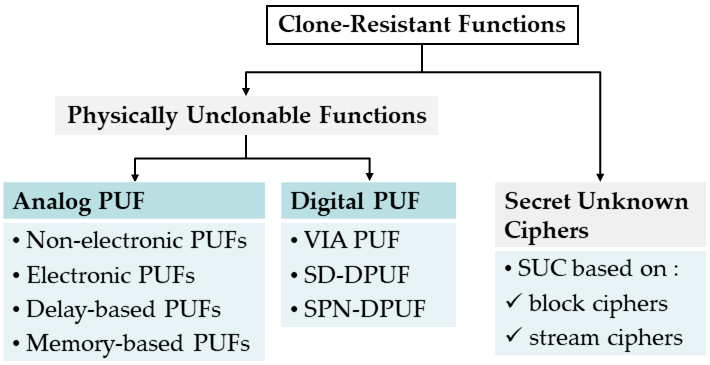}
\caption{Classification of clone-resistant functions.\label{fig1}}
\end{figure}

This section describes a summary of analog and digital clone-resistant functions. As depicted in Fig.~\ref{fig1}, clone-resistant functions can be categorized into (1) PUFs that can be either analog or digital, and (2) Secret Unknown Ciphers that are classified into SUCs based on random block ciphers and SUCs based on random stream ciphers.
The following sections provide basic design construction, properties and limitations for each of the described clone-resistant functions in Fig.~\ref{fig1}.

\subsection{Physical(ly) Unclonable Functions}
PUFs are primitives making use of intrinsic electronic, non-electronic, or physical devices’ properties to extract unique identity for each device. They are categorized into analog and digital PUFs.

\subsubsection{Analog PUFs or Mismatch-Based PUFs}
Many analog PUFs instances were proposed in the literature \cite{ref1}\cite{ref9} as described in Fig.~\ref{fig1}. Electronic, delay-based and memory-based PUFs deploy intrinsic properties of electronic devices to extract unique chips identities. Whereas, the construction and/or operation of non-electronic PUFs is inherently non-electronic, however, electronic circuits are used to process and store the PUF’s responses. Analog PUFs have two main downsides: inconsistency issues and their vulnerability to cloning attacks.
First, analog PUFs response spaces are noisy, requiring the use of FEs or HDA to stabilize their responses \cite{ref2}\cite{ref3}. FE generates and stores helper-data during the enrollment phase, which will be used, in the field, to reconstruct the original PUF response from a PUF noisy response. 
Second, PUFs are vulnerable to many attacks; modeling attacks represent a strong threat in cloning strong PUFs. D. Lim introduced the first attack to model an Arbiter-Based PUF \cite{ref10} and later on Majzoobi et al. analyzed linear and feed-forward PUF structures \cite{ref11}. Recently, Rührmair et al. demonstrated PUF modeling attacks on many PUFs by using machine learning techniques \cite{ref12}. Semi-invasive means have been used to reveal the state of memory-based PUFs \cite{ref13}. In \cite{ref14}, side channel attack was used to analyze PUFs architecture and fuzzy extractor implementations by deploying power analysis. Recent attack trends combine both side channel and modeling attacks \cite{ref15}. In \cite{ref16}, a hybrid attack is presented, combining side channel analysis and machine learning for attacking especially weak PUFs which prohibit attackers to observe their outputs.

\subsubsection{Digital PUFs or Physical-Based PUFs}
The random responses in physical-based PUFs are obtained from whether or not the conducting layers in a semiconductor are physically connected or not. Since these physical connections are not influenced by external factors such as temperature and supply voltage variations, physical-based PUFs can reach close to perfect reliability. Three physical-based PUFs were proposed in the literature:
\begin{itemize}
    \item \textbf{VIA PUF:} Vertical Interconnect Access (VIA) PUF was proposed in \cite{ref17}\cite{ref18}, and it uses the probability of via formation to generate unique and robust ICs’ responses. VIA PUF is a weak PUF having only one response.
    \item \textbf{SD-PUF:} Spliced Digital PUF (SD-PUF) \cite{ref19} takes advantage of the randomness from VLSI interconnect, namely the metal wires, that can be either connected or disconnected. In \cite{ref19}, the interconnect randomness is realized by intentionally positioning two interconnect layout line-ends close to each other, and due to mask variations, the generated masks will have mismatches. This Boolean connection (connected/disconnected) is called virtual connection. Such mismatch leads to uncertain connectivity status. SD-PUF combines multiple Digital PUFs (D-PUF) from multiple “building-chips”; D-PUF consists of N rows and M columns of unit cells. Each unit cell consists of a 2-input XOR gate where one of its inputs is connected to an input key bit and the other is connected to the strongly skewed-1 latch, which is connected to a virtual connection pin as source of the randomness. SD-PUF has multiple challenge-responses and hence it can be categorized under strong PUFs.
    \item \textbf{SPN-DPUF:} Substitution Permutation Network Digital PUF (SPN-DPUF) was proposed recently in \cite{ref35}. SPN-DPUF consists of three layers:  X-layer implementing a D-PUF as in \cite{ref19}, slayer and a player as in \cite{ref20}. SPN-DPUF is a strong digital PUF, it has additional hardware overhead compared to SD-PUF and has similar statistical properties as SD-PUF.
\end{itemize}
Digital PUFs or physical-based PUFs have consistent responses to some extent. However, they have two main limitations: (1) they can be used only for ASIC designs, and (2) the design assumptions cannot always be reached in practice, mainly because of the limited drawing resolution differences when shifting from wafer to wafer. 

\subsection{Secret Unknown Ciphers}
SUCs are digital clone-resistant functions with no instability issues as digital PUFs, whereas SUCs are implemented in System on Chip (SoC) FPGAs requiring no changes on the chip design. Also, SUCs can be implemented with about zero-cost when the FPGA resources are not totally used by the functional hardware designs. SUC-designs can deploy either random block ciphers or random stream ciphers as depicted in Fig.~\ref{fig1}. This paper focus on a SUC design based on random block ciphers.
\subsubsection{Definition}
SUC is a randomly and internally self-generated unknown and unpredictable cipher inside a chip, where users, manufacturers or operators have no access or influence on its creation process.\\
This work presents a lightweight involutive SUC based on random block cipher design. It can be defined as an involutive Pseudo Random Permutation ($iPRP$) as follows:

\begin{equation}
\begin{split}
SUC: \left \{ 0,1 \right \}^{n}\longrightarrow \left \{ 0,1 \right \}^{n}\\
X\xrightarrow{iPRP}Y
\end{split}
\end{equation}

\noindent Where: $SUC(SUC(X))=X$ for all $X\in \{0,1\}^{n}$.\par
Involutive SUCs are easier to implement in practice compared to non-involutive SUCs. Furthermore, SUC structures would have lower hardware/software complexities.

\subsubsection{Basic SUC Creation Concept}
\begin{figure}
\includegraphics[scale=0.3]{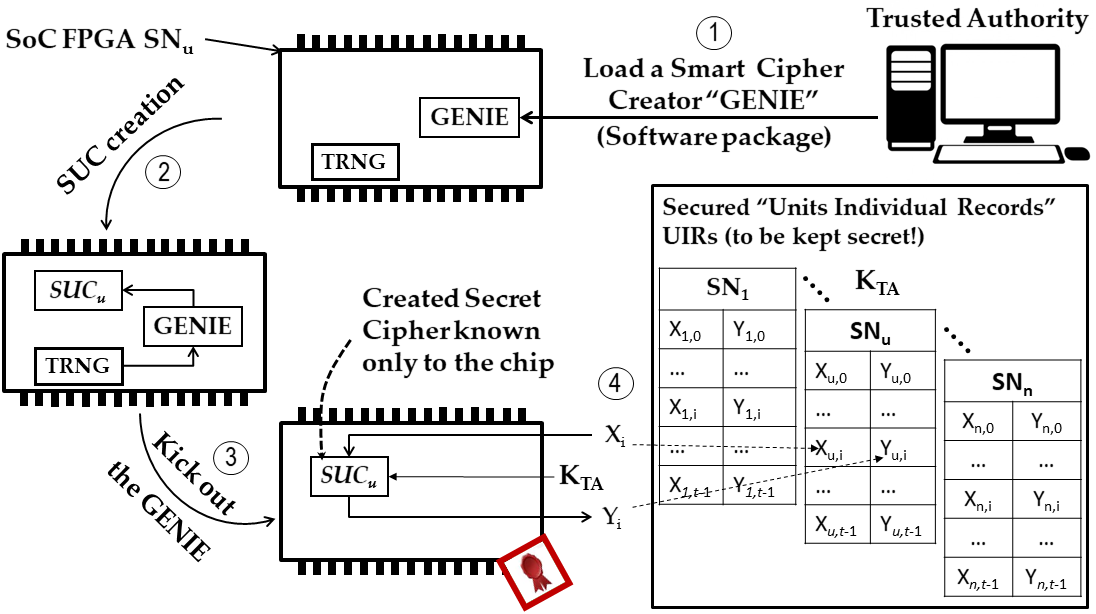}
\caption{The basic concept for creating SUC in SoC FPGAs environment, by a trusted authority in a trusted environment.\label{fig2}}
\end{figure}

Fig.~\ref{fig2} describes a possible scenario for embedding SUC in a System on Chip (SoC) non-volatile FPGA device. The personalization process proceeds as follows: 
\begin{itemize}
    \item \textit{Step 1:} A Trusted Authority (TA) uploads a software package called “GENIE” that contains an algorithm for creating internally secure random ciphers, and a set of cryptographically strong functions included to be used for randomly selecting the SUCs. The TA uploads the GENIE for a short time into each SoC FPGA unit to be used for just one time.
    \item \textit{Step 2:} After being loaded into the chip, the GENIE is triggered to create a permanent (non-volatile) and unpredictable random cipher. The cipher design components are completely randomly selected by deploying random bits from a True Random Number Generator (TRNG) within the chip.
    \item \textit{Step 3:} After completing the $SUC_u$ creation, the GENIE is completely deleted.
    \item \textit{Step 4:} by completing step 3, the SoC FPGA unit $u$ contains its unique and unpredictable $SUC_u$. TA then personalizes/enrolls the unit $u$ by challenging its $SUC_u$ with a plaintext challenge-set $\{X_{u,0}, X_{u,1} … X_{u,(t-1)}\}$ to get the corresponding ciphertext response-set $\{Y_{u,0}, Y_{u,1} … Y_{u,(t-1)}\}$. The two sets are stored securely as secret records in the Units Individual Records (UIR) labeled by the serial number of the device $SN_u$. UIRs are kept secret by TA. A secret key $K_{TA}$ may be added to the SUC design for multi-TA usage.     
\end{itemize}
The $X/Y$ pairs can be used later by TA to identify and authenticate devices. Notice that, the concept does not even allow TA or chip-manufacturer to create two entities with the same $SUC$.

\subsubsection{State of the Art of SUC Designs}
SUC designs are categorized into: 
\begin{itemize}
    \item \textit{SUC-based on Random Block Ciphers:} The first SUC design based on random block ciphers was proposed by the authors in \cite{ref21}\cite{ref22}. Authors present a novel concept for creating SUCs in non-volatile SoC FPGAs based on internal partial reconfiguration, this involves FPGA bitstream manipulation inside the chip. Two SUCs designs were proposed: an Involutive SUC (I-SUC) and a Non-Involutive SUC (NI-SUC) based on random involutive /non- involutive block ciphers. 
    \item \textit{SUC-based on Random Stream Ciphers:} In \cite{ref23}, an SUC design based on random stream ciphers was proposed, its key stream generator is based on a class of single-cycle T-functions with randomly selected parameters. In \cite{ref24}, authors propose a lightweight SUC based on a new family of stream ciphers. This SUC design is based on combining randomly selected $n$-bit nonlinear feedback shift registers (NLFSR). Each NLFSR’s feedback function is selected randomly from a set of feedback functions ensuring that the resulting NLFSR has a period of $2^n-1$.
\end{itemize}

\subsubsection{SUC-based Authentication}
In \cite{ref25}, a generic SUC-based authentication protocol was proposed, it also presents an efficient mechanism for Challenge-Response Pairs (CRPs) management. Those protocols are targeted for SUC-based on random block ciphers. In \cite{ref24}, authors present generic protocols for SUC-based on random stream ciphers. 
Recently, SUCs were proposed as building blocks in many security applications especially for devices authentication. In \cite{ref26}, SUC is proposed for special use to be embedded in vehicular electronic control units; a variety of automotive security applications were developed based on SUCs such as secure in-vehicle network, secure and private vehicle-to-vehicle, vehicle-to-roadside communications, and secure Over-The-Air (OTA) software update. The proposed OTA software update is based on building a chain of trust by deploying SUCs invertibility. A new concept for SUC-based secured e-coins circulations was proposed in \cite{ref27}, and an anonymous fair exchange e-commerce deploying SUC-enabled hardware tokens was proposed in \cite{ref28}.

\section{Novel Practical Concept for SUC Creation for Ultra-Low Latency Authentication}
\subsection{Motivation}
In \cite{ref22}, a novel concept allowing partial self-reconfiguration in future SoC FPGAs was proposed, it is based on internal partial bitstream manipulation. Authors present efficient approaches to embed this mechanism in SoC FPGAs with low-cost overhead. To create SUCs, it was proposed to deploy an SUC-design-template (SDT) where some mappings in the SUC design are selected randomly from a corresponding set. This operation is to be done internally inside the chip by the GENIE. Despite the novelty and efficiency of the proposed method in \cite{ref22}, it requires having full-knowledge about the configuration bitstream format, and hence it can be implemented efficiently even by the SoC FPGA manufacturer or when having access to the configuration bitstream format. 
\par
Thus, a practical concept for creating SUC is required when having no access or information about the configuration bitstream format, this constitutes the main purpose of this work. In the following, a new concept for creating random ciphers in contemporary SoC FPGAs is described. The realization of this concept is performed on SmartFusion®2 SoC FPGAs and it can be implemented similarly in all modern volatile and non-volatile SoC FPGAs.

\subsection{Novel SUC Creation Mechanism}
This section describes a novel SUC creation mechanism. As a requirement, the following components should be embedded within the targeted SoC FPGA to implement the proposed SUC creation mechanism:
\begin{itemize}
    \item Microcontroller for running the GENIE processes, it should also embed a sufficient non-volatile memory for storing parts of the GENIE.
    \item Cryptographic cores: mainly, a cipher such as AES, PUF with its FE/HDA (or pseudo-PUF) and a TRNG.
    \item An FPGA fabric with SRAM memory blocks
\end{itemize}
The proposed SUC creation mechanism proceeds in three steps:
\begin{itemize}
    \item SUC Design Template (SDT) creation
    \item One-Time SUC personalization
    \item SUC reinitialization
\end{itemize}
In the following, each step is described in details.

\subsubsection{SUC Design Template Creation}
\begin{figure}
\centering \includegraphics[scale=0.45]{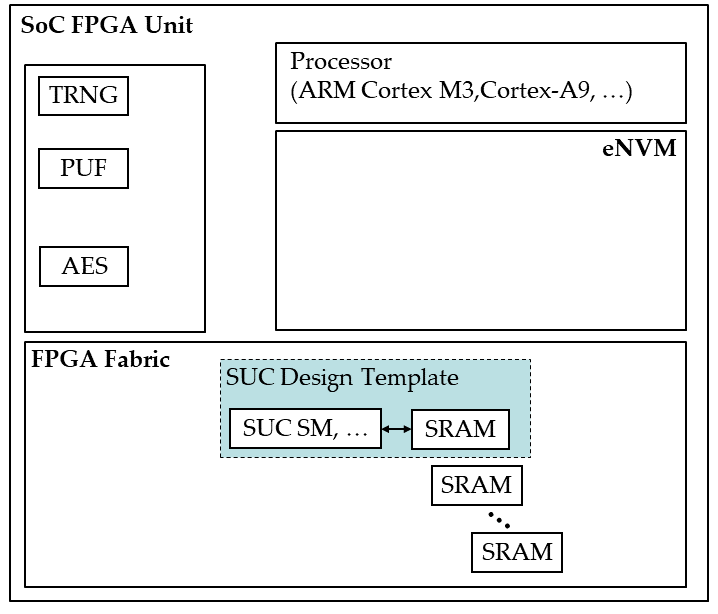}
\caption{SUC design template creation process.\label{fig3}}
\end{figure}

The first step to create an SUC is to design a cipher where some or all of its mappings or keys can be selected randomly, here we opt for SUC Design Template (SDT) concept proposed by A. Mars et al. in \cite{ref21}\cite{ref22}. In this work, the SDT makes use of some logic elements and SRAM blocks as described in Fig.~\ref{fig3}. The logic elements are used to implement SUC state machines (SUC SM), multiplexers, etc. Whereas the SRAM blocks are used to store the random mappings or keys loaded to it during reinitialization process. The SDT is an HDL design that can be compiled incrementally and added to the end product design as proposed in \cite{ref22}. The resulting configuration bitstream would be used to configure all SoC FPGAs equally where the area location of the SDT is fixed. The configuration bitstream contains also the software application that will run in the processor (ARM Cortex M3, Cortex-A9, …). The software application includes the GENIE Applications Programming Interfaces (APIs). It is also possible that each manufacturer locates the SDT in its preferable area location in the FPGA fabric. 

\subsubsection{One-Time Personalization Process}
\begin{figure}
\centering \includegraphics[scale=0.44]{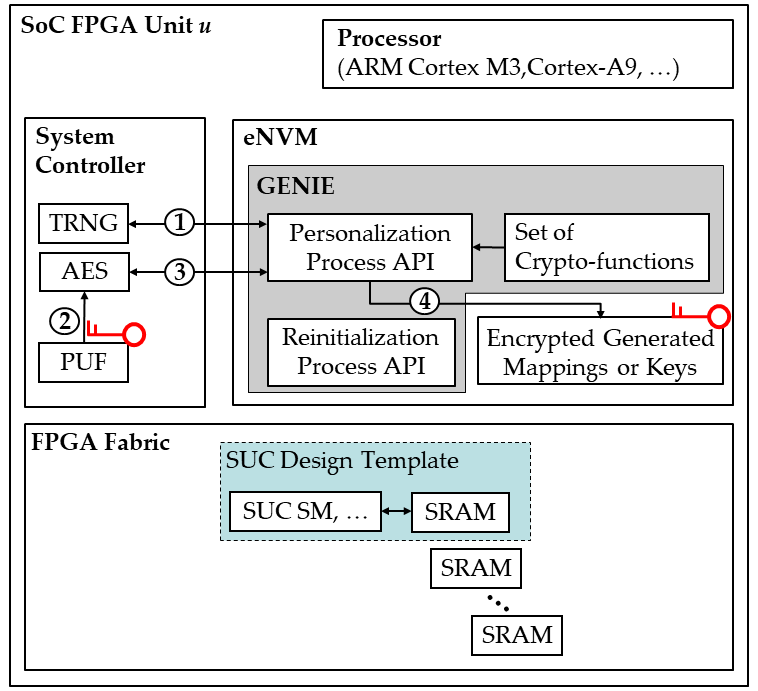}
\caption{One-time personalization process of SoC FPGA unit $u$.\label{fig4}}
\end{figure}

After finalizing the SDT creation process, each SoC FPGA embeds the same SDT in the FPGA fabric. The One-Time Personalization Process (OTPP) is to be accomplished by the GENIE, which is a software application that resides in the embedded Non-Volatile Memory (eNVM) as shown in Fig.~\ref{fig4}.
\par
The GENIE contains three components; first, a set of cryptographic mappings, which accommodate well selected mappings with good cryptographic properties that will be used during personalization process to create unique SUCs. Second, the reinitialization process API that will be described in the next section, and last the personalization process API.
\\
\noindent The OTPP proceeds as follows:
\begin{enumerate}
    \item OTPP gets random numbers from the TRNG and selects randomly a number of mappings from the set of cryptographic mappings. It is also possible to generate a set of random keys that can be used with the SUC design. 
    \item After that, the selected mappings should be encrypted with a standard cipher. Two key options for this cipher can be defined:
    \begin{itemize}
        \item For SoC FPGA devices embedding a PUF, the cipher is keyed by a memoryless key provided by the PUF core. 
        \item For SoC FPGAs without PUF, a pseudo-PUF key can be deployed as AES key.
    \end{itemize}
    \item A standard cipher within the SoC FPGA, such as AES, is keyed even by the memoryless key from a PUF or a pseudo-PUF key and used to encrypt the selected mappings in step (1). This protects the mappings from being disclosed to an adversary. 
    \item The encrypted mappings or keys are stored in the eNVM.
\end{enumerate}

After completing the personalization process, the personalization process API and the set of cryptographic functions should be erased from the eNVM.

\subsubsection{Reinitialization Process}
\begin{figure}
\includegraphics[scale=0.44]{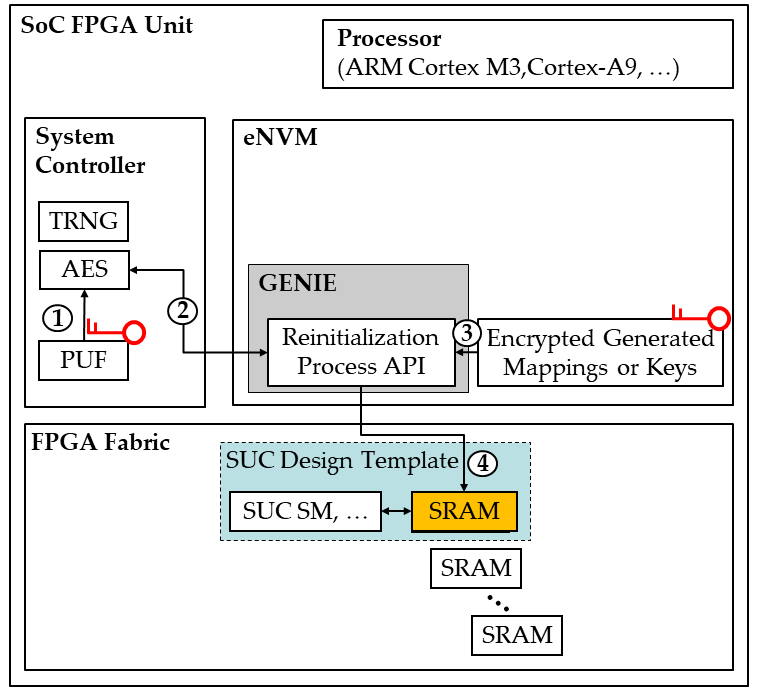}
\caption{Reinitialization process of SoC FPGA unit $u$.\label{fig5}}
\end{figure}

The SUC random mappings or keys are stored in a volatile memory (SRAM) block(s) in the FPGA fabric. After each power-on, reinitializing the SRAM block(s) is mandatory. This operation is accomplished by the Reinitialization Process (RP) API which decrypts the stored encrypted mappings or keys stored in the eNVM and loads the clear data to the SRAM block(s) as shown in Fig.~\ref{fig5}. The RP API should be kept stored in the eNVM. RP API proceeds as follows:

\begin{enumerate}
    \item Retrieve the memoryless PUF key or pseudo-PUF key 
    \item Read the encrypted mapping or keys from the eNVM
    \item Deploy the standard cipher (AES) within the SoC FPGA to decrypt the read data from the eNVM
    \item Load the clear mappings to the SRAM block (s) in the FPGA fabric.
\end{enumerate}
Notice that, the random ciphers are generated internally such that the random components selections depend on some random and unpredictable values generated by the TRNG. Hence, the randomly generated cipher is unknown to all other parties even the manufacturer. Here, it is assumed that the chip manufacturer does not fake the creation process by embedding some known random bits to be used.

\subsection{Implementation Use-Case}
The following section will describe an SDT based on substitution permutation network structure deploying 8-bit S-Boxes. The 8-bit S-Boxes are generated by using 4-bit S-Boxes as will be described in section IV.B. The set of cryptographic functions contains all Serpent-type 4-bit S-Boxes. The OTPP API deploys some random numbers from the TRNG to select a number of 4-bit S-Boxes from the set of cryptographic functions, then it generates the eight 8-bit S-Boxes, and stores them in an encrypted form in the eNVM. After each power-on, the RP API decrypts the stored encrypted eight 8-bit S-Boxes and loads the decrypted data to one large SRAM block that is used by the SDT. The proposed SDT in this work is coined as SRAM-SUC.

\section{SRAM-SUC DESIGN}
\subsection{Structure of SRAM-SUC}
\begin{figure}[!hb]
\noindent \includegraphics[width=\columnwidth]{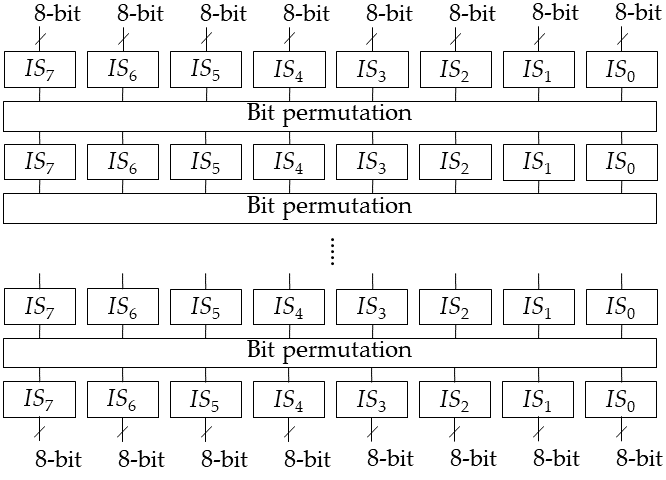}
\caption{SRAM-SUC: involution SDT using a class of 8-bit S-Boxes.\label{fig6}}
\end{figure}

The proposed concept for creating SUC in SoC FPGAs deploys SRAM to embed randomly selected mappings by the GENIE. For this purpose, the proposed SDT should have some mappings that can be embedded in the fabric SRAM blocks. S-Boxes are deployed in many block cipher designs, mostly to build confusion layers. They can be implemented even by using LUTs or by deploying memory blocks. By considering the possible configuration of SRAM blocks, many $n$-bit S-Boxes could be implemented efficiently, the design choice in this work is 8-bit S-Boxes and this will be sustained in section VI.B.
Fig.~\ref{fig6} describes the proposed 64 bits involutive block cipher deploying 8-bit S-Boxes coined as SRAM-SUC. All rounds use the same substitution layer (eight 8-bit S-Boxes) generated by randomly selecting them from the S-Boxes set that will be described in section IV.B. The diffusion layer is performed by using an involutive bit-permutation shown in section IV.C.  Notice that the last round has only a substitution layer to result with an involutive design.

\subsection{Generating a class of 8-bit S-Boxes from 4-bit S-Boxes}
\begin{figure}
\centering \includegraphics[scale=0.7]{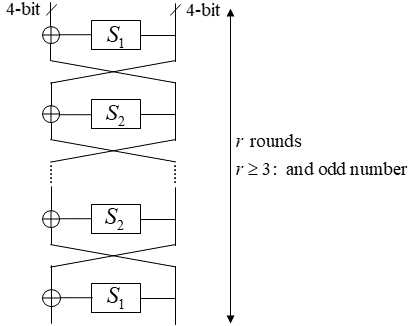}
\caption{Proposed design of 8-bit S-Boxes class from 4-bit S-Boxes.\label{fig7}}
\end{figure}

Involutive 8-bit S-Boxes can be generated, for instance, by using the inversion mapping in $GF(2^8)$. However, there exist only 30 irreducible polynomials of degree 8 in $GF(2^8)$ resulting with 30 possible involutive 8-bit S-Boxes.  
We propose a new methodology to generate a large class of involutive 8-bit S-Boxes by deploying the set of optimal 4-bit S-Boxes, as Feistel functions, as described in Fig.~\ref{fig7}. Involutive 8-bit S-Boxes are constructed by using an odd number of rounds $r$ of balanced Feistel network.

To have an involutive 8-bit S-Box, the design should be symmetric. The following conditions should be fulfilled:

\begin{equation}
    {{S}_{r-i-1}}={{S}_{i}}\,\,\,\,\text{with}\,\,\,\,\,i\in \left[ 0:\frac{r-1}{2}-1 \right]
\end{equation}

The number of 4-bit S-Boxes $S_i$ that can be selected randomly from the set of 4-bit S-Boxes is: 
\begin{equation}
    \frac{r+1}{2}
\end{equation}

Notice that, the same number is possible to select randomly when using $r'=r+1$ rounds (i.e. $r'$ is even).

\subsubsection{Class of Optimal 4-bit S-Boxes}
Let $ S:\mathbb{F}_{2}^{4}\to \mathbb{F}_{2}^{4}$  be a 4-bit S-Box. Let $Lin(S)$ and $Diff(S)$ denote the linearity and the differential resistance of $S$, respectively. An optimal 4-bit S-Box fulfills the following conditions:  

\begin{enumerate}
    \item $S$ is a bijection
    \item $Lin(S)=8$
    \item $Diff(S)=4$
\end{enumerate}

A Serpent-type S-Box \cite{ref29} fulfills the previous conditions. in addition to that, any one-bit input difference causes at least two bits output difference. According to \cite{ref29} there exist  $2,211,840=2^{21}$ such S-Boxes.
All S-Boxes in this class have an average differential and linear probabilities $p=2^{-2}$.
The following section provides a theoretical analysis of the cryptographic properties for the resulting large class of 8-bit S-Boxes generated from the class of Serpent-type 4-bit S-Boxes.  

\subsubsection{Cryptographic Properties of the Class of 8-bit S-Boxes}
In the design of Fig.~\ref{fig7}, a class of 8-bit S-Boxes can be generated by selecting randomly $(r+1)/2$ 4-bit S-Boxes from the set of Serpent-type 4-bit S-Boxes, where $r$ is an odd number of balanced Feistel rounds as in Fig.~\ref{fig7}.
The following theorem is adapted from \cite{ref30} and used to characterize the cryptographic properties of the resulting class of 8-bit S-Boxes.

\noindent \textbf{Theorem.} For the 8-bit S-Box design in Fig.~\ref{fig7} with $r\geqslant 3$, if the average differential probability (respectively the average linear probability) of the bijective 4-bit S-Boxes $S_i$ $0\leq i \leq \frac{r-1}{2}$  is smaller than $p$, then the resulting 8-bit S-Box has an average differential probability (respectively, average linear probability) smaller than $p^2$.\\
Optimal 4-bit S-Boxes have an average differential and linear probability $p=2^{-2}$.

\subsection{Involutive bit permutation}
In order to use the same design for both encryption and decryption operations, the diffusion layer should also be an involution. In \cite{ref31}, authors describe one involution player with good cryptographic properties that we formulate as follows: 
\begin{equation}
    \left [ IS_{i} \right ]_{j}=\left [ IS_{j} \right ]_{i}
\end{equation}

Where output bit $j$ of the involutive S-Box $i$ ($IS_i$) is fed to the input bit $i$ of the involutive S-Box $j$ ($IS_j$) in the next round. Note that, this linear transformation can only be used in block ciphers with block length $N$ using $M$ equal S-Boxes, each having $n$-bit inputs such that $M=N/n$.

\subsection{Cardinality of SRAM-SUC}
\subsubsection{Cardinality of the Class of 8-bit S-Boxes}
The number of optimal 4-bit S-Boxes is  $\left | S \right |\approx 2^{21}$ \cite{ref29}. The design of 8-bit S-Boxes (Fig.~\ref{fig7}) uses $(r+1)/2$ randomly selected 4-bit S-Boxes from the set of Serpent-type S-Boxes, where it is possible to use the same 4-bit S-Box in all rounds.\\
Hence, the cardinality of all possible resulting 8-bit S-Boxes is:
\begin{equation}
\left| \varsigma  \right|={{\left| S \right|}^{\frac{r+1}{2}}}
\end{equation}

\subsubsection{Cardinality of SRAM-SUC}
SRAM-SUC deploys eight 8-bit S-Boxes as described in Fig.~\ref{fig6}. Each 8-bit S-Box is generated randomly and can be seen as a selection from the class of cardinality $\left| \varsigma  \right|$. Hence, the cardinality of SRAM-SUC is:

\begin{equation}
\begin{split}
  \left| SRAM\text{-}SUC \right| &={{\left| \varsigma  \right|}^{8}}={{\left( {{\left| S \right|}^{\frac{r+1}{2}}} \right)}^{8}}\\
  &\approx {{({{({{2}^{21}})}^{\frac{r+1}{2}}})}^{8}}={{2}^{84r+84}}
\end{split}
\end{equation}
For $r=13$, this results with $|SRAM-SUC|=2^{1176}$.

\section{STATISTICAL ANALYSIS OF SRAM-SUC}
\subsection{Avalanche effect}
\begin{figure}
\centering \includegraphics[width=\columnwidth]{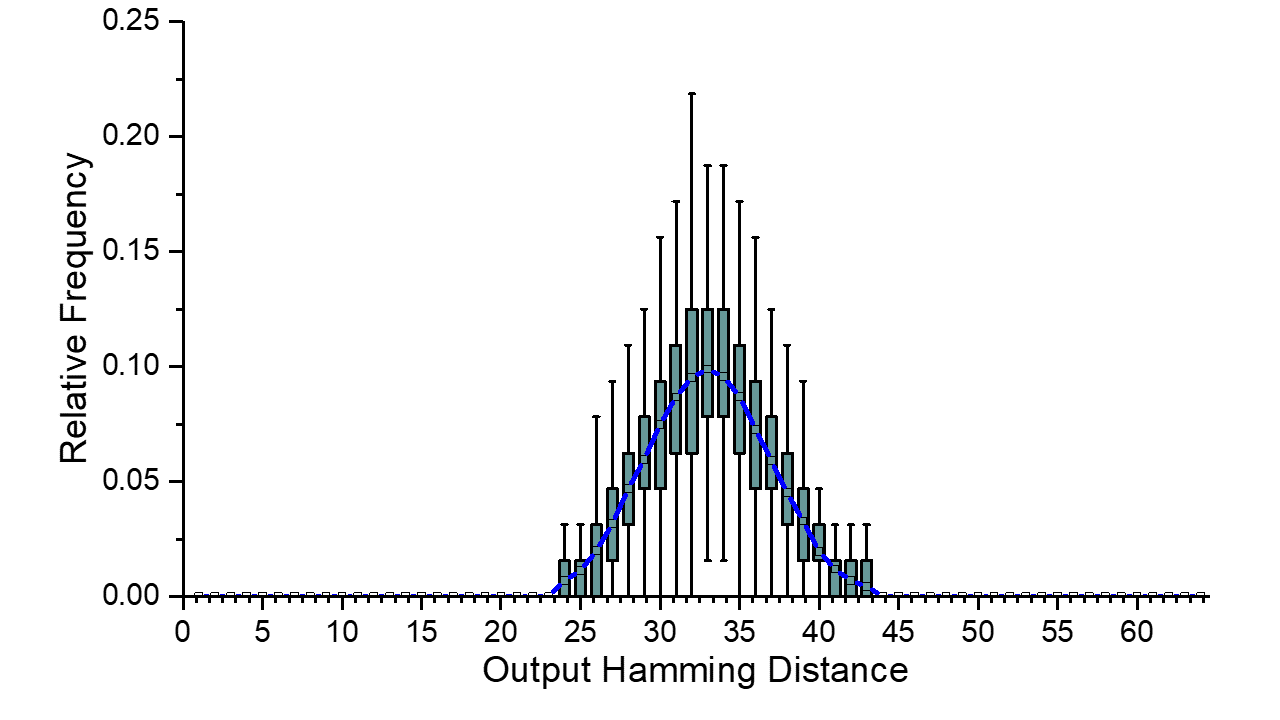}
\caption{Distribution of the output hamming distance when testing the avalanche effect for SRAM-SUC.\label{fig8}}
\end{figure}

This section presents statistical analysis results for SRAM-SUC.  According to the presented mechanism for creating SUCs, the GENIE selects randomly $4(r+1)$ 4-bit S-Boxes from a set of optimal 4-bit S-Boxes and uses them to generate randomly eight 8-bit S-Boxes. Hence, to check possible avalanche characteristics of SRAM-SUCs, SUCs deploying different S-boxes should be considered. Simulations results are for $r=3$, which consists the lowest number of rounds used to generate 8-bit S-Boxes.\\
\noindent This experiment tests the avalanche characteristics as follows:
\begin{itemize}
    \item A set of 1000 optimal 4-bit S-Boxes is used to construct 1000 8-bit S-Boxes.
    \item 1000 SRAM-SUCs are generated where each SRAM-SUC uses one S-Box from the selected set.
    \item We used 100 random numbers to evaluate the avalanche characteristic. For each number, the number of output bit changes is measured when flipping each of its bits.
\end{itemize}
In the ideal case, the expected number of bit changes should have a binomial distribution with the peak at the half of the number of output bits (32 in our case). Fig.~\ref{fig8} shows an almost perfect binomial distribution for SRAM-SUC.

\subsection{Avalanche Characteristic of SRAM-SUC}
\begin{figure}
\centering
\includegraphics[width=\columnwidth]{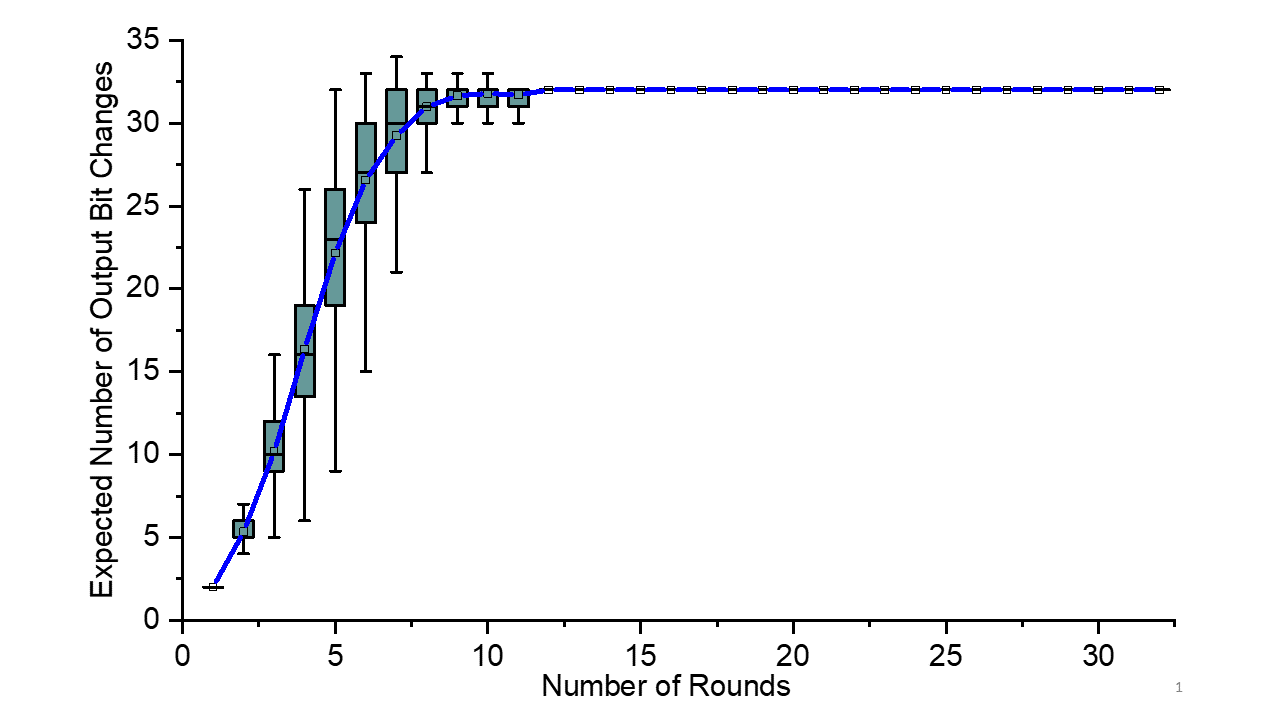}
\caption{Expected number of output bit changes in function of the number of rounds.\label{fig9}}
\end{figure}

To evaluate the effect of number of rounds on the avalanche characteristic, we conducted similar experiment as the previous one with different number of rounds ranging from 1 to 32. Fig.~\ref{fig9} shows the ranges of output bit changes in function of the number of rounds, the mean of the number of output bit changes is plotted in blue line.

\section{Implementation Methodology}
\subsection{General Description of the SUC Core}
\begin{figure*}
\centering \includegraphics[width=2.0\columnwidth]{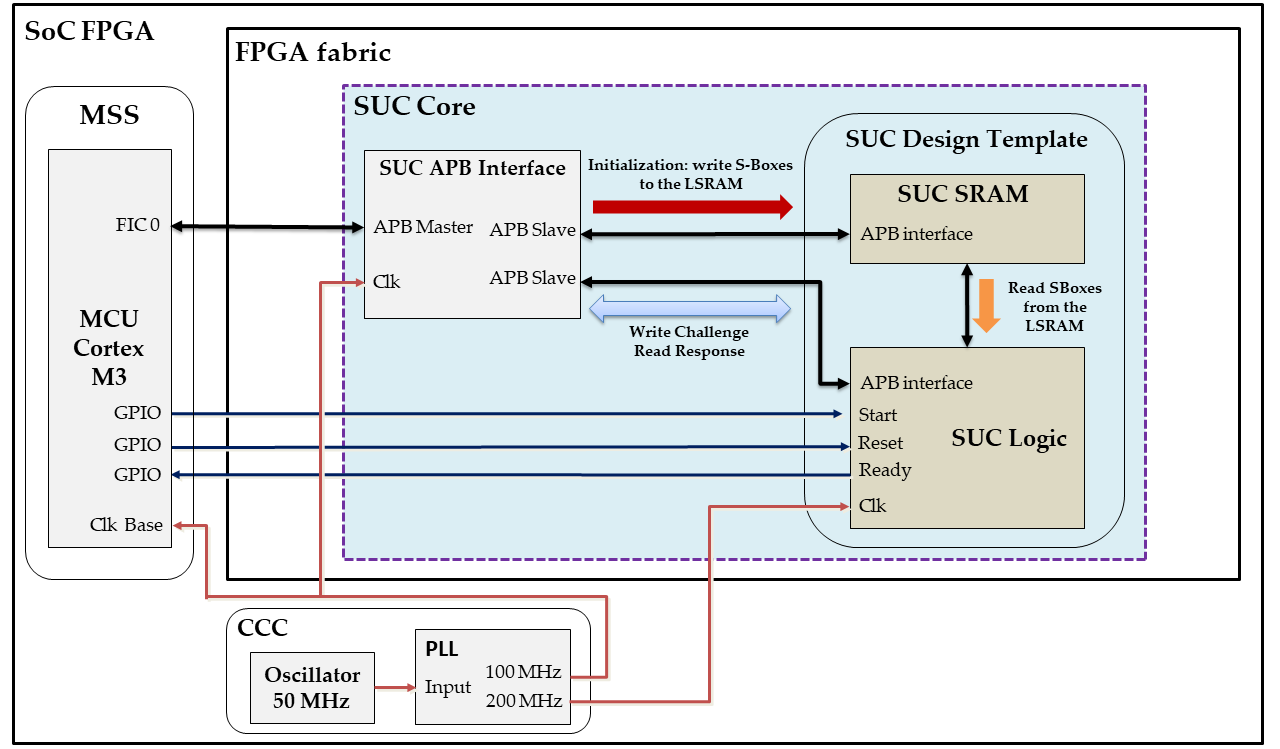}
\caption{General structure of the SUC core.\label{fig10}}
\end{figure*}

Fig.~\ref{fig10} describes the general structure of the SUC core. The SUC core consists of the SRAM-SUC (SDT) with an APB slave interface to receive/send challenges/responses to the MSS.  The LSRAM block is connected to the MSS via an APB slave interface for reinitialization process. The MSS can only write on the LSRAM and is not physically allowed to read.\par
The control signals of the SUC core are connected directly to the GPIOs of the MSS. The APB interfaces are used as a data bus for the following tasks:
\begin{enumerate}
    \item Write all the involutive S-Boxes (256 Byte x 8 S-Boxes) from the MSS to the LSRAM block during the reinitialization process.
    \item Send a challenge (64-bit) from the MSS to the SUC core.
    \item Receive a response (64-bit) from the SUC core to the MSS.
\end{enumerate}
The main oscillator (50MHz) is used to generate all the required clock frequencies via the PLL component. The MSS and APB interfaces are working with 100MHz, while the SRAM-SUC is clocked by 200MHz.

\subsection{Implementation of the SRAM-SUC}
\subsubsection{Implementation of 8-bit S-Boxes}
Modern FPGAs embed multiple hard-blocks such as SRAM and DSP blocks. These blocks can be accessed through the fabric routing architecture.\par
\begin{figure}
\includegraphics[width=\columnwidth]{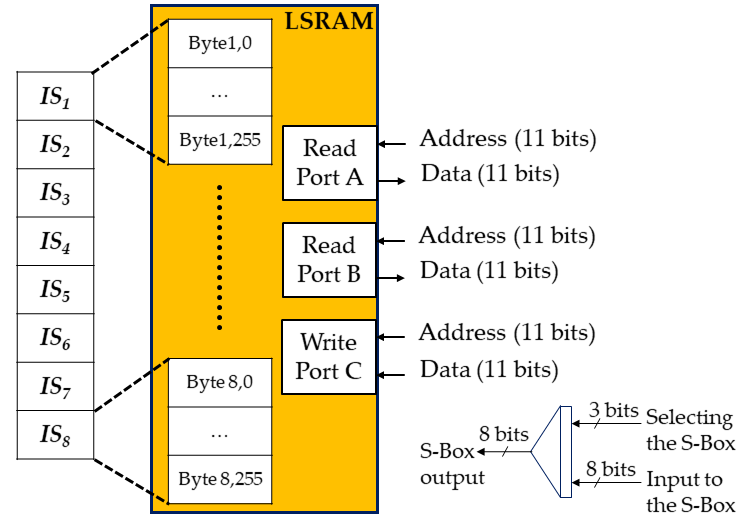}
\caption{Implementation of the eight 8-bit S-Boxes in one LSRAM block.\label{fig11}}
\end{figure}

SmartFusion®2 SoC FPGAs embed large SRAM (LSRAM) blocks, each can store up to 18,432 bits. Each LSRAM block can be configured in any of the following $depth\times width$ combinations: $512\times36$, $512\times32$, $1k\times18$, $1k\times16$, $2k\times9$, $2k\times8$, $4k\times4$, $8k\times2$, or $16k\times1$.Each 8-bit S-Box requires 256 bytes to store the S-Box output while its inputs will be provided as the access address. Consequently, the configuration $2k\times8$ is deployed to implement all the eight 8-bit S-Boxes as shown in Fig.~\ref{fig11}. The 11-bit access address are composed of 3 bits for the S-Box address ($IS_1$ to $IS_8$) and 8-bit as inputs to the selected S-Box. Note that each LSRAM block contains two read ports that will be deployed to provide 16 bits for each read rather than 8 bits. Hence, 4 cycles are required to generate the 64 bits outputs of the substitution layer.

\subsubsection{SRAM-SUC State Machine}
The SRAM-SUC state machine consist of three states: NOP, RUN, and READY. The SUC remains in the NOP state till having a trigger signal from the MSS to start encryption/decryption operation. During the RUN state, the SUC perform encryption/decryption operation and moves to the READY state when the response is ready. 

\subsubsection{Hardware Implementation of the SUC Core}
\begin{figure*}
\centering \includegraphics[width=2.0\columnwidth]{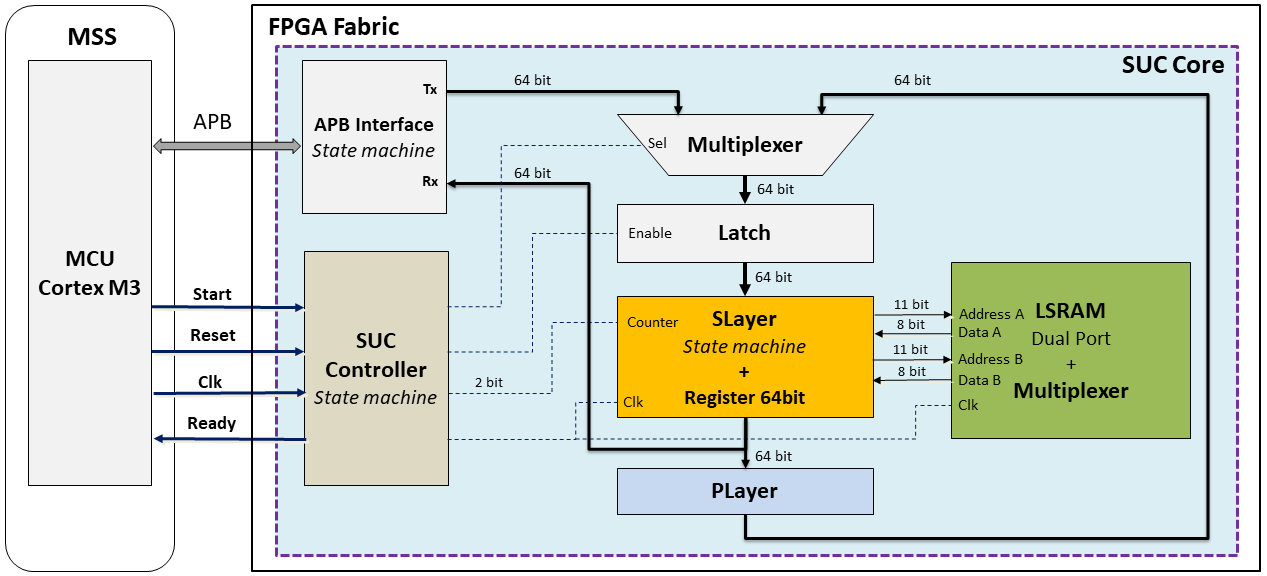}
\caption{Description of the implementation method of the SUC Core.\label{fig12}}
\end{figure*}

The implemented SUC core in the FPGA fabric consists of the SRAM-SUC (SDT) together with an APB interface as described in Fig.~\ref{fig12}. Following is a description of each SUC Core component:
\begin{enumerate}
    \item APB interface: It is used as main connection bus between the MSS and the FPGA fabric to send and receive challenge-response pairs between the MSS and the SUC.  
    \item Multiplexer (64bit): It is used to select between the input data from the APB interface and the output data from the PLayer. The default state allows to receive the challenge from the MSS via the APB interface.
    \item Latch: It consists of a 64-bit register with an enable control input from the SUC controller. The latch is used to store the input data that will be provided to the Slayer when receiving the order from the SUC controller.
    \item SLayer: It consist of a state machine that reads the corresponding involutive S-Boxes from the LSRAM. It has also a 64-bit register to save the output data of the SLayer.
    \item Player: A 64-bit permutation for the output data of the SLayer. Each input bit is connected, via a layer, to a different bit position of the multiplexer.
    \item SUC controller: it is a state machine that sends control signals to the other SUC core components and receive control signals from the MSS. 
\end{enumerate}
The MSS sends a plaintext to the SUC core via the APB interface. The maximum data width of the APB interface is     32-bit. Therefore, a state machine is used to handle the total input data 64-bit for the SUC core. \par
The SUC controller is triggered by a high level of the start input (start=’1’). By default, the multiplexer provides the 64-bit output from the APB interface (Tx) to the latch (select=’0’). At the second clock cycle, the SUC controller makes select=’1’ which stays in high level till the end of the “run state”. The SUC controller triggers the latch by putting enable=’1’. After that, the latch will provide its input data to the Slayer state machine. 
The Slayer state machine processes each two 8-bit data blocks from the latch separately. Each 8-bit data block constitutes an input to one 8-bit S-Box. The Slayer state machine should get the corresponding 8-bit S-Box output of each 8-bit input data block from the LSRAM. The LSRAM block is configured as 2048x8 bit cells with two read ports: A and B. Each read port has an input address of 11-bits to access all cells in the LSRAM. The three most significant bits are used to select one of the 8-bit S-Boxes, while the other 8 low significant bits are used as data block inputs to a selected 8-bit S-Box. Each read port (A and B) has an output data of 8-bit providing the output of two 8-bit S-Boxes in one cycle. The 16-bit (2x 8-bit) are stored directly in the output register of the Slayer. This process is repeated four times to provide the 64-bit output of the Slayer.
The output register of the Slayer state machine has two roles: firstly, during the “run state”, it provides the input data to the Player. Secondly, on the “ready state”, the APB interface gets the output of this register as the final response from the SRAM-SUC and sends it directly to the MSS.
\par
The Player (see Section IV.C) is implemented by using only connection layers between the output register (64-bit) of the Slayer and the multiplexer input, which is connected to the Latch starting from the second clock cycle.

\subsubsection{Performance of SRAM-SUC}
\begin{figure*}
\centering \includegraphics [width=2.0\columnwidth]{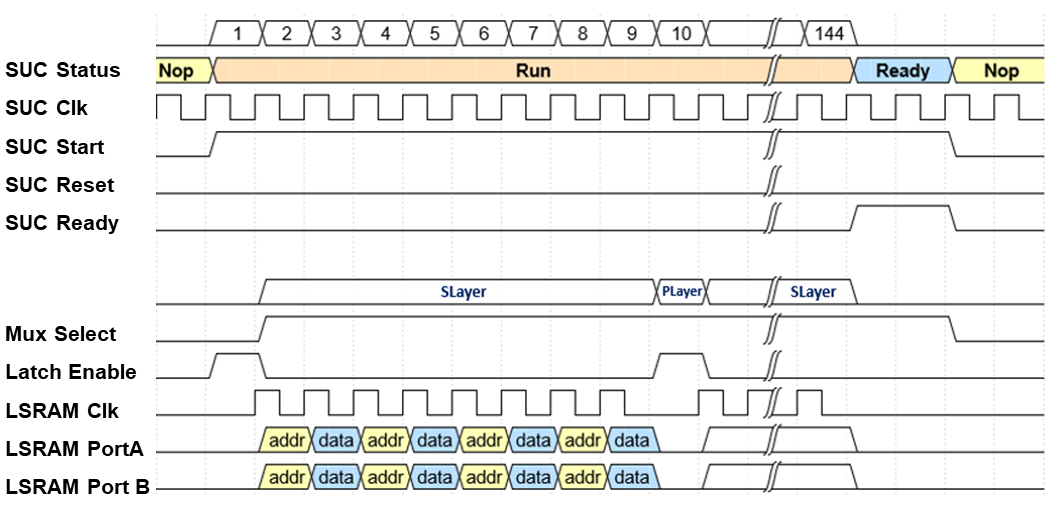}
\caption{Hardware performance of the SUC design template (SRAM-SUC).\label{fig13}}
\end{figure*}

The SDT requires 144 cycles to complete a full encryption or decryption for 64-bit data with 15 rounds as shown in Fig.~\ref{fig13}. 
The SUC remains in the NOP state until that it receives a high level of the start signal from the MSS. When start=’1’, the latch is activated with a positive edge of the signal “enable” to store the input data from the multiplexer, which is, by default, configured to get the data from the APB interface (select=’0’). In the next positive edge of the clock signal Clk, the latch’s enable signal returns to ‘0’ and the multiplexer control signal select=’1’ allowing direct connection between the Player and the Latch. At this stage, the data is ready to be processed by the Slayer which takes 8 clock cycles to generate the corresponding output of the Slayer; in each clock cycle, the Slayer state machine provides two access addresses in port A and B to read the data (2x 8-bit) from the corresponding two LSRAM cells in the next clock cycle. This process is repeated four times to generate the 64-bit Slayer output which is stored in a 64-bit register. At the clock cycle number 10, the latch is activated to store the first-round data. The full round is repeated 15 times while, in the last round, the final SUC response is taken directly from the Slayer output register (no Player at the last round). After the clock cycle 144, the SUC moves to the ready state and stays in this state till that start=’0’.  

\section{Implementation results}
\subsection{Hardware Complexity and Performance of SRAM-SUC}
\subsubsection{Hardware Complexity of SRAM-SUC}
This section presents the implementation results of the proposed SDT (SRAM-SUC) in the FPGA fabric.\\
TABLE~\ref{table1} presents the hardware resources usage by the SUC core, which contains the SDT together with the SUC APB interface. The SDT includes the SUC logic and SUC SRAM, which contains one LSRAM with its interface to the SUC logic and the APB slave for reinitialization process. The complete SUC core deploys 1 LSRAM block, 191 LUTs, and 208 DFFs. In terms of clusters, this requires 18 logical clusters in addition to 1 LSRAM block with its interface clusters that are automatically used.
\begin{table}

\caption{Hardware complexity of SRAM-SUC}
\label{table1}

\begin{center}
\renewcommand{\arraystretch}{1.5}
\begin{tabular}{|l|l|l|l|l|l|l|l|l|l|}
\hline
\multicolumn{2}{|l|}{\multirow{2}{*}{\textbf{Module}}}                               & \multicolumn{4}{l|}{\textbf{Resources usage}} & \multicolumn{4}{c|}{\begin{tabular}[c]{@{}c@{}}\textbf{\% of resources usage}\\ \textbf{in M2S005}\end{tabular}} \\ \cline{3-10} 
\multicolumn{2}{|l|}{}                                                      & \rotatebox{90}{LUT}  & \rotatebox{90}{DFF} & \rotatebox{90}{LSRAM Block}  & \rotatebox{90}{eNVM (KB)} & \rotatebox{90}{LUT}  & \rotatebox{90}{DFF} & \rotatebox{90}{LSRAM Block}  & \rotatebox{90}{eNVM (KB)}    \\ \hline

\multicolumn{2}{|l|}{MSS}                                                   & 0    & 0   & 0           & 2         & 0        & 0        & 0               & 1.56         \\ \hline
\multicolumn{2}{|l|}{CoreAPB3}                                              & 38   & 0   & 0           & 0         & 0.62     & 0        & 0               & 0            \\ \hline
\multirow{2}{*}{\begin{tabular}[c]{@{}l@{}}SRAM\\ SUC\end{tabular}} & SRAM  & 27   & 4   & 1           & 0         & 0.44     & 0.06     & 10              & 0            \\ \cline{2-10} 
                                                                    & Logic & 126  & 204 & 0           & 0         & 2.07     & 3.36     & 0               & 0            \\ \hline
\multicolumn{2}{|l|}{Total}                                                 & 191  & 208 & 1           & 2         & 3.15     & 3.43     & 10              & 1.56         \\ \hline
\end{tabular}%
\end{center}
\end{table}

\subsubsection{Hardware Performance of SRAM-SUC}
TABLE~\ref{table2} describes the execution time of the SRAM-SUC in the FPGA fabric. For one encryption/decryption, the SRAM-SUC requires 2.88 $\mu s$ when running at a frequency of 50 MHz and 0.72  $\mu s$ when clocked with 200 MHz.

\begin{table}
\renewcommand{\arraystretch}{1.5}
\caption{Execution time of SRAM-SUC}
\label{table2}
\begin{center}
\begin{tabular}{|l|l|l|l|l|}
\hline
\textbf{Operations}                                                                        & \begin{tabular}[c]{@{}l@{}}\textbf{Number}\\ \textbf{of Cycles}\end{tabular} & \begin{tabular}[c]{@{}l@{}}\textbf{SDT Clk}\\ {[}MHz{]}\end{tabular} & \begin{tabular}[c]{@{}l@{}}\textbf{Data Size}\\ {[}bit{]}\end{tabular} & \begin{tabular}[c]{@{}l@{}}\textbf{Time}\\ {[}$\mu s${]}\end{tabular} \\ \hline
\multirow{2}{*}{\begin{tabular}[c]{@{}l@{}}Encryption/\\ Decryption\end{tabular}} & \multirow{2}{*}{144}                                       & 50MHz                                                       & \multirow{2}{*}{64}                                           & 2.88                                                    \\ \cline{3-3} \cline{5-5} 
                                                                                  &                                                            & 200MHz                                                      &                                                               & 0.72                                                    \\ \hline
\end{tabular}

\end{center}
\end{table}

\subsection{Software GENIE}
The software GENIE consists of two APIs: 
\begin{itemize}
    \item One-Time Personalization Process API
    \item Reinitialization Process API
\end{itemize}
In the following, we present the software performance of both one-time personalization and reinitialization processes APIs. 

\subsubsection{Software Performance of the OTPP}
This section describes the software performance of the OTPP.\par
\noindent \textbf{Theorem.} Let $r$ be the number of Feistel network rounds used to generate 8-bit S-Boxes as in Fig.~\ref{fig7}. Let $|S|$ represents the number of optimal 4-bit S-Boxes, which constitute the class of cryptographic mappings stored with the GENIE package. The required number of random bytes from the TRNG is:
\begin{equation}
    {{\kappa }_{TRNG}}=4\times \left\lceil {{\log }_{2}}\left( \left| S \right| \right) \right\rceil \times \left( r+1 \right)\text{ Bytes}
\end{equation}

\noindent \textbf{Proof.} The number of randomly selected 4-bit S-Boxes to generate each 8-bit S-Box is $(r+1)/2$. Hence, the GENIE selects $4\times(r+1)$ 4-bit S-Boxes to generate all 8-bit S-Boxes.\\
The number of all optimal 4-bit S-Boxes is $|S|$, hence $\left\lceil {{\log }_{2}}\left( \left| S \right| \right) \right\rceil$ random bits are required to select randomly one 4-bit S-Box out of this set. The theorem follows.
\par
The time complexity induced by triggering the TRNG to generate ${{\kappa }_{TRNG}}$ bytes is:

\begin{equation}
    {{\tau }_{TRNG}}={{\tau }_{1}}\times {{\kappa }_{TRNG}}+{{\tau }_{2}}
\end{equation}

Where ${{\tau }_{1}}=4.78125 \mu s$ and ${{\tau }_{2}}=388 \mu s$. These parameters are derived from experimental results using the random number generator embedded in SmartFusion®2 SoC FPGA.\\
To generate an 8-bit S-Box, the OTTP executes   rounds, a linear fitting of the implementation results shows that the time complexity to generate an 8-bit S-Box is:
\begin{equation}
    {{\tau }_{r}}={{\tau }_{3}}\times r+{{\tau }_{4}}
\end{equation}
Where ${{\tau }_{3}}=1.63 \mu s$ and ${{\tau }_{4}}=0.07 \mu s$.\\
After generating each 8-bit S-Box, the OTPP encrypts the resulting S-Box and then stores the result in the eNVM. \\

Let ${{\tau }_{e}}$ denotes the encryption time complexity, ${{\tau }_{PUF}}$ denotes the required time to retrieve the PUF key, and   denotes the required time to store all the encrypted S-Box in the eNVM. Hence, the time complexity of the OTPP is:
\begin{equation}
    {{\tau }_{OTPP}}={{\tau }_{TRNG}}+{{\tau }_{PUF}}+8\left( {{\tau }_{r}}+{{\tau }_{e}} \right)+{{\tau }_{envm}}
\end{equation}
It can be expressed in simpler form as:
\begin{equation*}
{{\tau }_{OTPP}}={{k}_{1}}\times \left\lceil {{\log }_{2}}\left( \left| S \right| \right) \right\rceil \left( r+1 \right)+{{k}_{2}}\times r+{{k}_{3}}
\end{equation*}

\noindent where ${{k}_{1}}=4{{\tau }_{1}}\approx \text{19}\text{.125}\,\mu s\,\,\text{;}\,\,{{k}_{2}}=8{{\tau }_{3}}\approx 13.04\mu s$;\\
${{k}_{3}}={{\tau }_{2}}+{{\tau }_{PUF}}+8\left( {{\tau }_{4}}+{{\tau }_{e}} \right)+{{\tau }_{envm}}\approx 647 ms$\\

\noindent For $r=3$ and $|S|=256$, we have: ${{\tau }_{OTPP}}\approx 647.679 ms$. \\
Table III shows the experimental results of the OTPP execution time for the same parameters values $r=3$ and $|S|=256$.
\\

\begin{table}
\renewcommand{\arraystretch}{1.5}
\caption{Software performance of the personalization process API}
\label{table3}
\begin{center}
\begin{tabular}{|c|c|c|c|}
\hline
\textbf{\begin{tabular}[c]{@{}c@{}}Personalization \\ Processes\end{tabular}} & \textbf{\begin{tabular}[c]{@{}c@{}}MSS Clk \\ (MHz)\end{tabular}} & \textbf{\begin{tabular}[c]{@{}c@{}}Data Size\\ (Byte)\end{tabular}} & \textbf{\begin{tabular}[c]{@{}c@{}}Time\\ ($ms$)\end{tabular}} \\ \hline
\begin{tabular}[c]{@{}c@{}}Random Number \\ Generator\end{tabular}            & \multirow{5}{*}{100MHz}                                           & 16                                                                  & 0.464                                                        \\ \cline{1-1} \cline{3-4} 
Retrieve PUF key                                                              &                                                                   & 16                                                                  & 30                                                           \\ \cline{1-1} \cline{3-4} 
\begin{tabular}[c]{@{}c@{}}4bit to 8bit Sbox \\ Generator\end{tabular}        &                                                                   & 2048                                                                & \multirow{2}{*}{22}                                          \\ \cline{1-1} \cline{3-3}
AES256 Encryption                                                             &                                                                   & 2048                                                                &                                                              \\ \cline{1-1} \cline{3-4} 
Writing to eNVM                                                               &                                                                   & 2048                                                                & 596                                                          \\ \hline
\multicolumn{3}{|c|}{Total Personalization Time}                                                                                                                                                                        & 648.464                                                      \\ \hline
\end{tabular}

\end{center}
\end{table}

\begin{figure}
\centering \includegraphics [width=\columnwidth]{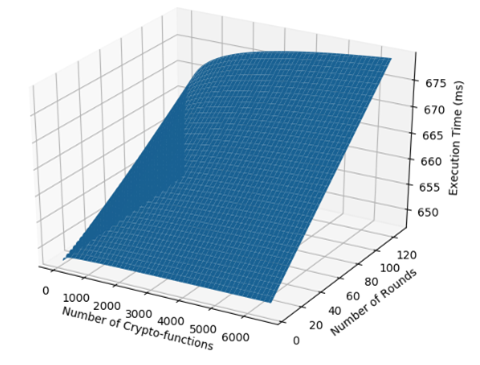}
\caption{Execution time of OTPP in function of the number of rounds $r$ and the crypto-functions set cardinality $|S|$.\label{fig14}}
\end{figure}

Fig.~\ref{fig14} describes the execution time of the OTPP, according to the derived model in the previous theorem, in function of the number of rounds $r$ and the S-Boxes cardinality $|S|$.

\subsubsection{Software Performance of the Reinitialization Process}
After each power-on, the SoC FPGA runs the reinitialization process to retrieve its SUC. This API reads first the encrypted eight 8-bit S-Boxes from the eNVM, then it retrieves the PUF key, and uses it to decrypt the 2048 bytes, and loads them to the LSRAM block. TABLE IV presents the software performance results. The total reinitialization time is 51 ms and it is constant.
\begin{table}[H]
\renewcommand{\arraystretch}{1.5}
\caption{Software performance of the reinitialization process API}
\label{table4}
\centering \begin{tabular}{|c|c|c|c|}
\hline
\textbf{\begin{tabular}[c]{@{}c@{}}Personalization \\ Processes\end{tabular}} & \textbf{\begin{tabular}[c]{@{}c@{}}MSS Clk \\ (MHz)\end{tabular}} & \textbf{\begin{tabular}[c]{@{}c@{}}Data Size\\ (Byte)\end{tabular}} & \textbf{\begin{tabular}[c]{@{}c@{}}Time\\ ($ms$)\end{tabular}} \\ \hline
Reading from eNVM                                                             & \multirow{4}{*}{100MHz}                                           & 2048                                                                & 1.33                                                         \\ \cline{1-1} \cline{3-4} 
Retrieve PUF key                                                              &                                                                   & 16                                                                  & 30                                                           \\ \cline{1-1} \cline{3-4} 
AES256 Decryption                                                             &                                                                   & 2048                                                                & 18                                                           \\ \cline{1-1} \cline{3-4} 
Writing to LSRAM                                                              &                                                                   & 2048                                                                & 1.77                                                         \\ \hline
\multicolumn{3}{|c|}{Total Reinitialization Time}                                                                                                                                                                       & 51                                                           \\ \hline
\end{tabular}
\end{table}

\section{Comparative Analysis}
To make an accurate performance comparison between our proposal and the existing PUF-based authentication mechanisms, all implementations are performed on SmartFusion2 M2S150 SoC FPGA embedding SRAM PUF and Elliptic Curve Cryptography (ECC) used also for challenge-response purpose.

\subsection{SRAM PUF Functionalities in Microsemi FPGAs}
SRAM PUF is a widely deployed PUF instance; a commercial SRAM PUF is manufactured by Intrinsic ID \cite{ref32}. It is embedded in many chips such as Microsemi large SmartFusion2 and IGLOO2 FPGAs devices, Intel Stratix 10, and NXP LPC5500 series, etc. SRAM PUF can be used for memoryless key generation and storage, authentication, and it can also provide random seeds. 
\subsubsection{Memoryless Key Storage}
Microsemi ‘S’ grade SmartFusion2 SoC FPGAs (-060, -090 and -150 devices) embed Quiddikey IP core from Intrinsic ID with 2KB SRAM \cite{ref33}. The deployment of SRAM PUF for key generation requires three steps \cite{ref33}\cite{ref34}: 
\begin{itemize}
    \item \textit{Enrollment process:} it is used to generate an activation code (AC) of 1192 byte based on the startup state of the SRAM. The AC is stored in the eNVM for future use in generating user keys. 
    \item \textit{Key Code Generation:} an activation code and intrinsic or extrinsic keys are used to generate key codes. A user can enroll many intrinsic or extrinsic keys.
    \item \textit{Key Reconstruction:} an activation code and a key code are utilized to reconstruct the intrinsic or extrinsic key.
\end{itemize}
The enrollment is necessary to be done only one time. Users can retrieve an intrinsic/extrinsic key by only running the key reconstruction process.

\subsubsection{Current Device Challenge-Response Mechanism in SmartFusion2 and IGLOO2}
The large devices of ‘S’ grade SmartFusion2 and IGLOO2 (-60, -90, -150) embed an authentication mechanism based on Elliptic-Curve Cryptography (ECC). The manufacturer adds a random unique 384-bit key ($K_{ECC}$) to each device during the key provisioning step, it is claimed that this key is not recorded or re-constructible by the manufacturer. This key is protected using the SRAM PUF, i.e., it is enrolled as an extrinsic key. \par
As described in the previous section, the enrollment of $K_{ECC}$ is required to be done once, Quiddikey IP will generate a corresponding key code $KC_{ECC}$ and stores it in the eNVM. Quiddikey IP can reconstruct $K_{ECC}$ using $KC_{ECC}$ and the Activation Code (AC). \par
\begin{figure}
\centering \includegraphics[scale=0.43]{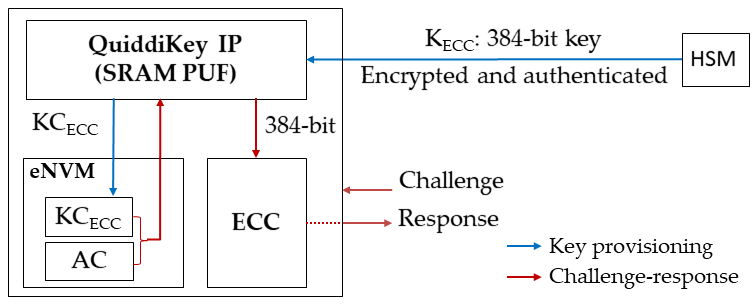}
\caption{Challenge-response mechanism in Microsemi SmartFusion2 SoC FPGAs, ‘S’ grade, and large devices ($\ge 60K$).\label{fig15}}
\end{figure}

To enroll a device, several challenges are generated, and the corresponding responses are recorded. To authenticate a device, a user checks a response of a same recorded challenge as proof that the device is the same. Fig.~\ref{fig15}  describes the challenge-response mechanism in ‘S’ grade large SmartFusion2 and IGLOO2 devices.

In the following, we provide an accurate performance comparison between our proposed SRAM-SUC, Quiddikey IP SRAM PUF embedded in SmartFusion2 SoC FPGAs, and the challenge-response mechanism based on ECC recommended to be used in SmartFusion2 and IGLOO2 ‘S’ grade large devices. 
TABLE V presents the performance comparison results. A basic retrieving of an intrinsic or extrinsic key requires 30$ms$, while the challenge-response mechanism based on ECC requires 33.8$ms$ for the first challenge-response after power-up and 22.5$ms$ for the other challenge-response operations. The SRAM-SUC requires only 0.72$\mu s$ when clocked at 200MHz. Hence, the execution time of SRAM-SUC is tremendously faster with a factor of 41666 and 30250 than Quiddikey IP and ECC CR service, respectively. 

\begin{table}
\renewcommand{\arraystretch}{1.5}
\caption{Performance comparison between SRAM PUF and SRAM-SUC}
\label{table5}
\begin{center}
\begin{tabular}{|c|c|c|c|}
\hline
\textbf{Function}                                                                & \multicolumn{2}{c|}{\textbf{Components}}                                              & \textbf{Latency ($ms$)} \\ \hline
\multirow{3}{*}{\begin{tabular}[c]{@{}c@{}}Quiddikey IP\\ SRAM PUF\end{tabular}} & \multirow{2}{*}{\begin{tabular}[c]{@{}c@{}}One\\ Time\end{tabular}} & Activation Code & 600                   \\ \cline{3-4} 
                                                                                 &                                                                     & Enroll Key      & 646                   \\ \cline{2-4} 
                                                                                 & \multicolumn{2}{c|}{Retrieve Key (C-R mode)}                                          & 30                    \\ \hline
\multirow{3}{*}{SRAM-SUC}                                                        & \multicolumn{2}{c|}{One-Time Personalization}                                         & 664                   \\ \cline{2-4} 
                                                                                 & \multicolumn{2}{c|}{Reinitialization (each power-up)}                                 & 51                    \\ \cline{2-4} 
                                                                                 & \multicolumn{2}{c|}{SRAM-SUC (C-R mode)}                                              & 0.00072               \\ \hline
\multicolumn{3}{|c|}{ECC CR Service}                                                                                                                                     & 33.8/22.5             \\ \hline
\end{tabular}

\end{center}
\end{table}

\subsubsection{Enrollment Process of an SUC-enabled Device by a TTP}
TABLE VI shows the measured time values of the enrollment process with different numbers of CRPs. The communication between the trusted authority and the SUC is deploying basic CAN bus 2.0 with baud rate 100 Kbps.
\\
\begin{table}[H]
\renewcommand{\arraystretch}{1.5}
\caption{Timing analysis of the enrollment process}
\label{table6}
\begin{adjustbox}{width=\columnwidth,center}
\begin{tabular}{|c|c|c|c|c|c|}
\hline
\textbf{\begin{tabular}[c]{@{}c@{}}Number\\ of Pairs\end{tabular}} & \textbf{\begin{tabular}[c]{@{}c@{}}CAN\\ Speed\end{tabular}} & \textbf{\begin{tabular}[c]{@{}c@{}}MSS Frequency\\ (MHz)\end{tabular}} & \textbf{\begin{tabular}[c]{@{}c@{}}SUC Clk\\ (MHz)\end{tabular}} & \textbf{\begin{tabular}[c]{@{}c@{}}Data Size\\ (Byte)\end{tabular}} & \textbf{\begin{tabular}[c]{@{}c@{}}Time\\ (s)\end{tabular}} \\ \hline
16                                                                 & \multirow{4}{*}{100Kbps}                                     & \multirow{4}{*}{100MHz}                                                & \multirow{4}{*}{200MHz}                                          & 256                                                                 & 1.2                                                         \\ \cline{1-1} \cline{5-6} 
32                                                                 &                                                              &                                                                        &                                                                  & 512                                                                 & 2.34                                                        \\ \cline{1-1} \cline{5-6} 
1024                                                               &                                                              &                                                                        &                                                                  & 16384                                                               & 72.2                                                        \\ \cline{1-1} \cline{5-6} 
2048                                                               &                                                              &                                                                        &                                                                  & 32768                                                               & 144                                                         \\ \hline
\end{tabular}
\end{adjustbox}
\end{table}

\section{Conclusion}
This paper has presented an ultra-low latency robust digital PUF, coined as SRAM-SUC, allowing to achieve devices authentication in URLLC applications. The creation of SRAM-SUC is accomplished using a novel mechanism for creating secure random and unpredictable ciphers in SoC FPGAs. The concept realization was done on SmartFusion®2 SoC FPGA and, similarly, it can be realized in other SoC FPGAs. The proposed SRAM-SUC structure is designed to make use of existing FPGA resources especially memory blocks (uSRAM and LSRAM). This concept can be used for different types of SUC design templates such that the randomly selected mapping or part of them should be stored in the SRAM inside the FPGA fabric. Many SUC designs with $n$-bit S-Boxes can be implemented since the SRAM blocks offer different $length\times width$ configurations.

\bibliography{refs}
\bibliographystyle{IEEEtran}
\end{document}